\documentclass[useAMS]{mn2e}
\usepackage{epsfig,graphics}
\usepackage{amsmath}

\title[SED role in IDV of Blazars]
{Peak of spectral energy distribution play an important role in intra-day variability of Blazars?}

\author[Gupta et al.]{Alok C.\ Gupta$^{1,2}$\thanks{Email: acgupta30@gmail.com}\thanks{CAS Visiting Fellow, CAS PIFI Visiting
Scientist},
Nibedita\ Kalita$^{2,3}$\thanks{Email: nibeditaklt1@gmail.com},
Haritma\ Gaur$^{1}$\thanks{Email: haritma@shao.ac.cn},
Kalpana\ Duorah$^{3}$ \\
\\
\\
$^1$Key Laboratory for Research in Galaxies and Cosmology, Shanghai Astronomical Observatory, Chinese Academy of Sciences, \\
~~80 Nandan Road, Shanghai 200030, China \\
$^2$Aryabhatta Research Institute of Observational Sciences (ARIES), Manora Peak, Nainital, 263 002, India \\
$^3$Department of Physics, Gauhati University , Guwahati,781014, India}

\begin{document}
\date{Accepted \dots Received \dots; in original form \dots}

\maketitle

\label{firstpage}

\begin{abstract}
Blazars can be divided into two sub-classes namely high energy and low energy peaked blazars. In spectral energy 
distribution, the first synchrotron hump of the former class peaks in UV/X-rays and in IR/optical bands for the 
latter class. The peak of the spectral energy distribution seems to be responsible for variability properties of 
these classes of blazars in X-ray and optical bands. Since, in low energy peaked blazars, the X-ray bands lies 
well below the synchrotron hump, one expects that the highest energy electrons available for the synchrotron 
emission would have 
%retarded effect 
slower effect of variability on X-ray intra-day timescale. In this paper, by taking the advantage of a sample of 12 low 
energy peaked blazars with total 50 observations from XMM$-$Newton since its launch, we confirm that this class 
is less variable in X-ray bands. We found that out of 50 observational light curves, genuine intra-day variability 
is present in only two of light curves i.e 4\%. Similar results we obtained from our earlier optical intra-day 
variability studies of high energy peaked blazars where out of 144 light curves, only genuine intra-day variability 
was detected in 6 light curves i.e $\sim$ 4\%.
 Since, X-ray bands lie below the peak of the spectral energy distribution of LSPs where inverse Compton mechanism 
is dominating rather than synchrotron radiation at the peak of the optical band, leads to slower variability
%the retarded effect of electrons 
in the X-ray bands. Hence, reducing their intra-day variability in X-ray bands as compared to the variability in optical bands.
\end{abstract}


\section{Introduction}

A small sub-class of radio-loud active galactic nuclei (AGN) is known as blazars. BL Lac objects and flat spectrum 
radio quasars (FSRQs) collectively known as blazars. Optical spectrum of BL Lac objects are featureless i.e. absence 
of prominent emission or absorption lines, whereas FSRQs show prominent emission lines. The common properties of blazars 
include large amplitude violent flux variation in complete electromagnetic (EM) spectrum, high and variable polarization 
from radio to optical bands, core-dominated radio morphology, and emission being predominantly nonthermal. Blazars emit 
relativistic charged particle jets which pointed close (at angles $\leq$ 10$^{0}$) to our line of sight (e.g., Urry \& Padovani 
1995) and it causes the observed emission to be relativistically beamed.

Since blazars emit radiation in the complete EM spectrum, these are among ideal objects to study their multi-wavelength spectral 
energy distribution (SED). The SED of blazars show two well defined broad spectral components (Mukherjee et al. 1997). Based
on the location of these SED peaks, blazars are further classified into low energy peaked blazars (LBLs) and high energy peaked 
blazars (HBLs). In LBLs the first SED component peaks in radio to optical while the second component peaks at GeV energies, and 
in HBLs the first component peaks in UV/X-rays while the second component peaks at TeV energies (Padovani \& Giommi 1995). Recently 
blazars classification is made on synchrotron peak frequency and divided into three sub-classes i.e low synchrotron peaked (LSPs),
intermediate synchrotron peaked (ISPs) and high energy peaked blazars (HSPs) (Abdo et al. 2010a). Basically LSPs and ISPs collectively 
belong to LBLs class. 

\begin{table*}

{\bf Table 1.} Observation log of XMM-Newton X-ray data for low energy peaked blazars$^{*}$
\small

\begin{tabular}{lccclcclr} \hline
Blazar Name    & $\alpha_{2000.0}$& $\delta_{2000.0}$             & redshift &Blazar & Date of Obs. & Obs. ID    & Window     & GTI$^{b}$(s) \\
               &                  &                               &   $z$    &Class & yyyy.mm.dd   &            & Mode$^{a}$ &              \\\hline

TXS 0106$+$612 & 01h09m46.3s      & $+$61$^{0}$33$^{'}$30$^{''}$  & 0.783    & ~LSP & 2010.02.09   & 0652410201 & ~~FF         & 12034  \\
PKS 0235$+$164 & 02h38m38.9s      & $+$16$^{0}$36$^{'}$59$^{''}$  & 0.94     & ~LSP & 2002.02.10     & 0110990101 & ~~FF         & 16847  \\
               &                  &                               &          & & 2004.01.18        & 0206740101 & ~~SW         & 29671  \\
               &                  &                               &          & & 2004.08.02        & 0206740501 & ~~SW         & 11471  \\
               &                  &                               &         &  & 2005.01.28        & 0206740701 & ~~SW         & 16270  \\
PKS 0426$-$380 & 04h28m40.4s      & $-$37$^{0}$56$^{'}$20$^{''}$  & 1.11    & ~LSP  & 2012.02.11     & 0674330201 & ~~EFF        & 20745  \\
PKS 0528$+$134 & 05h30m56.4s      & $+$13$^{0}$31$^{'}$55$^{''}$  & 2.06    & ~LSP & 2009.09.11     & 0600121601 & ~~FF         & 25672  \\
PKS 0537$-$286 & 05h39m54.3s      & $-$28$^{0}$39$^{'}$56$^{''}$  & 3.104   & ~LSP  & 2000.03.19     & 0114090101 & ~~FF         & 18852  \\
               &                  &                               &         &  & 2005.03.20        & 0206350101 & ~~FF         & 80138  \\
S5 0716$+$714  & 07h21m53.4s      & $+$71$^{0}$20$^{'}$36$^{''}$  &  0.31   & ~ISP  & 2007.09.24     & 0502271401 & ~~SW         & 71624  \\
4C 71.07       & 08h41m24.3s      & $+$70$^{0}$53$^{'}$42$^{''}$  & 2.172   & ~LSP  & 2001.04.12     & 0112620101 & ~~FF         & 33480  \\
OJ 287         & 08h54m48.9s      & $+$20$^{0}$06$^{'}$31$^{''}$  & 0.3056  & ~ISP  & 2005.04.12     & 0300480201 & ~~LW         & 13192  \\
               &                  &                               &         &  & 2005.11.03        & 0300480301 & ~~LW         & 39074  \\
               &                  &                               &         &  & 2006.11.17        & 0401060201 & ~~LW         & 44972  \\
               &                  &                               &         &  & 2008.04.22        & 0502630201 & ~~LW         & 53568  \\
               &                  &                               &         &  & 2011.10.15        & 0679380701 & ~~LW         & 21669  \\
3C 279         & 12h56m11.1s      & $-$05$^{0}$47$^{'}$22$^{''}$  & 0.5362  & ~LSP  & 2009.01.21     & 0556030101 & ~~FF         & 25235  \\
               &                  &                               &         &  & 2011.01.18        & 0651610101 & ~~SW         & 125487 \\
BL Lac         & 22h02m43.3s      & $+$42$^{0}$16$^{'}$40$^{''}$  & 0.0686  & ~ISP  & 2007.07.10     & 0501660201 & ~~SW         & 18478  \\
               &                  &                               &         &  & 2007.12.05        & 0501660301 & ~~SW         & 19272  \\
               &                  &                               &         &  & 2008.01.08        & 0501660401 & ~~SW         & 22371  \\
3C 454.3       & 22h53m57.7s      & $+$16$^{0}$08$^{'}$54$^{''}$  & 0.859   & ~LSP  & 2006.07.02     & 0401700201 & ~~SW         & 15972  \\
               &                  &                               &         &  & 2007.05.23        & 0401700401 & ~~SW         & 2995   \\
               &                  &                               &         &  & 2006.12.18        & 0401700501 & ~~SW         & 15021  \\
               &                  &                               &         &  & 2007.05.31        & 0401700601 & ~~SW         & 28971  \\\hline
\end{tabular}     \\
$^{*}$ Observation log for the LSP 3C 273 is given in Table 1 of our paper Kalita et al. (2015). \\ 
$^{a}$ Extended Full Frame = EFF, Full Frame = FF, Large Window = LW, Small Window = SW \\
$^{b}$ GTI = Good Time Interval \\
\end{table*}

\begin{figure*}
 \includegraphics[width=185mm]{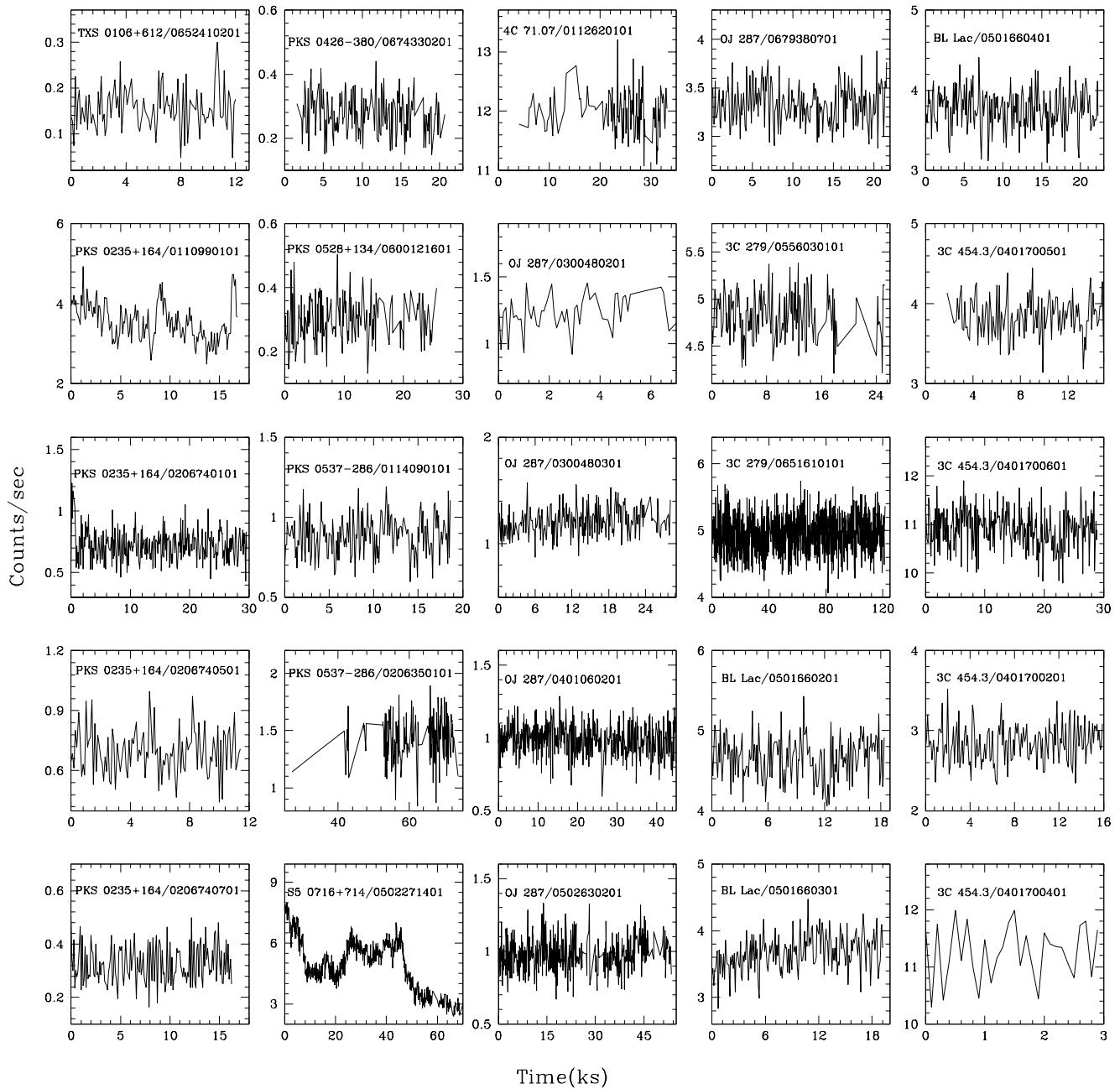}
 \caption{X$-$ray light curves of the low energy peaked blazars. In each panel blazar name / observation ID is given. 
Light curves are generated with 100s binning. For light curves of 3C 273 see Fig. 1 of Kalita et al. (2015).}
 \label{f1}
\end{figure*}

Variation in the blazar flux of the order of a few hundredth to tenths over a time scale of few minutes to less than a 
day is called intraday variability (IDV) (Wagner \& Witzel 1995). Variability timescales of weeks to few months is commonly 
known as short term variability (STV) and timescales of months to years is called long term variability (LTV) (Gupta et al. 
2004). 

The most puzzling flux variation in blazars are those which are happening on IDV timescales. The variability mechanisms in 
the blazars on IDV timescales are widely accepted that it is associated with the instabilities and irregularities in the 
jet flow. Hence, in order to understand IDV, it is very important to understand the fine details of jet formation. Astronomers 
seeks to image jet formation using very long baseline radio interferometry (VLBI) but it suffers severe lack of sufficient 
angular resolution. An alternative method to measure fine structure of jets i.e. to search for IDV with the fastest possible 
time sampled light curves (LCs).

The first genuine optical IDV in blazar was reported by Miller et al. (1989). Since then extensive search for optical IDV in 
large number of blazars were done (e.g. Heidt \& Wagner 1996; Montagni et al. 2006; Gupta et al. 2008, 2012; Agarwal et al. 2015; 
Gaur et al. 2015; and references therein). In a detailed statistical analysis of optical IDV of blazars, (Gupta \& Joshi 2005) 
reported that if a blazar continuously observed for less than 6 hours and more than 6 hours, the chances of detecting IDV is 
$\sim$ 60 -- 65\% and 80--85\%, respectively. Till 2005, most of the known blazars belong to LSPs and ISPs class and so the 
report by Gupta \& Joshi (2005) was true only for LSPs and ISPs. Till 2005, there were only 6 HBLs or HSPs (Mrk 421, Mrk 501, 
1ES 1426+428, 1ES 1959+658, PKS 2155$-$304, 1ES 2344+514) were known, and their studies were mainly focused in X-ray and 
$\gamma-$ray bands (e.g. Edelson et al. 2001; Zhang et al. 2002; Giebels et al. 2002; Krawczynski et al. 2002; Aharonian et al. 
2002, 2005; Konopelko et al. 2003; Massaro et al. 2004; Daniel et al. 2005; and references therein). Thanks to the revolution 
in $\gamma-$ray astronomy due to HESS, MAGIC, Fermi, VERITAS, etc. which lead to a very rapid increase in detection of new HSPs 
(e.g. Abdo et al. 2010b, Nolan et al. 2012, Acero et al. 2015). 

The new sample of HSPs gave us an opportunity to see the optical IDVs of HSPs and compare its properties with optical IDVs of 
LSPs. We started a dedicated project to search for optical IDV in HSPs and after doing 62 nights of IDV observations of HSPs 
which gave us 144 LCs (41 in B band, 62 in R band, and 41 in B$-$R color) of five HSPs (Mrk 421, 1ES 1426+428, 1ES 1553+113, 
1ES 1959+650, and 1ES 2344+514). Interestingly, we found that, 4 HSPs did not show any IDV (Gaur et al. 2012a, 2012b, 2012c), 
but only one HSP 1ES 1426+428 for which we have the least observations have shown IDV in 6 LCs out of 8 LCs (Gaur et al. 2012c). 
Our this pilot project gave us 6 IDV LCs out of 144 LCs searched for IDV i.e $\sim$ 4\% LCs have shown IDV. We explained it by 
density inhomogeneities and bends in the bases of the jets by Kelvin--Helmholtz instabilities (Romero et al. 1999). We gave an 
alternative explanation i.e. since in HSPs, the optical band lies below the SED peak, hence, we should see changes in the efficiency 
of acceleration of, and/or in the rates at which energy is radiated by, the highest energy electrons available for synchrotron 
emission would have a more retarded effect on optical variability in HSPs (Gaur et al. 2012b). 
In LSPs, the optical band is dominated by highest energy electrons emitting synchrotron radiation and probably the X-ray emission 
is dominated by the comparatively lower energy electrons emitting the inverse Compton radiation, hence their X-ray variability is less
pronounced than optical variability. If SED peak is really responsible 
for IDV properties, then we suspected that X--ray IDV LCs in LSPs should not show any IDV at all or show on rare occasions. 
With this motivation, here we present the X-ray IDV study of almost complete sample of 10 LSPs and 2 ISPs observed by XMM-Newton 
since its launch and we found that the LSPs show very less IDV 2 out of 50 LC i.e. 4\% in X-ray bands. We have reported above the 
similar finding for HSPs in optical bands. 

The paper is structured as follows. In Section 2, we discuss XMM--Newton archival data and analysis. Section 3 reports our results 
and section 4 contains discussion. 

\section{XMM-Newton Archival Data and Analysis}

\subsection{Observations and Data Analysis Technique}

XMM--Newton satellite is one of the most efficient X-ray observatory on board pointing X-ray sources, since 2000,
from its day of launch. Since then, the satellite's contribution in scientific community (X-ray astronomy) is enormous,
and this is not different for blazar studies. The satellite detect X-ray photon between the energy range 150 eV to
15 keV.  XMM-Newton gives one of the best time resolution, and the longest pointing of any AGN. Such data are extremely
useful for studying IDV of blazars.

Our sample contains all total 12 blazars (10 LSPs and 2 ISPs). 
Most of the sources in our sample are well known blazars and are extensively studied with different ground and space
based telescopes. But a few of them are recently come into focus of studies. XMM--Newton has observed all of these 
sources at least once in its on board period. In our study we use the EPIC/pn observations due to its higher reliability 
over EPIC/mos for very bright objects (Str{\"u}der et al.\ 2001). We excluded few observations, due to their high 
background brightness and few are excluded due to bad image quality. The observations we use in our analysis are listed 
in Table 1, with all important and necessary information. In the Table 1 we have listed as a whole, total 25 EPIC/pn 
observations excluding for the blazar 3C 273. For 3C 273 the observation log is given in Table 1 of (Kalita et al. 2015).

For data reduction we followed the standard method used for XMM--Newton data, using the {\it SCIENCE ANALYSIS SOFTWARE
(SAS)} version 12.0.1. Here we take the data within the energy range 0.3 keV to 10 keV (Kalita et al. 2015). Though we
restrict the high energy range of data train up to 10 keV, to avoid the flaring effect (mainly coming from Sun), still 
we found a few light curves get affected by this. The highly affected light curves by flaring are corrected by removing 
the affected portion from the light curves. 

Event files generated using the standard {\it SAS} routine {\it epchain} and later the source image was generated combining
both the single and double events $(PATTERN \leq 4)$, excluding the events that are at the edge of a CCD $(FLAG $=$ 0)$.
The source area for extracting the source light curves varies from 30 -- 40 arcsec depending on the position of the source
on the CCD chip. Finally the background subtracted light curves are generated with a time bin of 100 seconds, using the
task {\it epiclccorr}.

Pile up was found for only one source, 3C 273. We tackle this problem using an annuls area instead of circular 
one for both the source and the background regions with inner circular source region of 10 arcsec and outer region of 
40 arcsec as average for all observations affected by pile up. More detail of this process is mentioned in (Kalita et al. 
2015). Few observations excluding 3C 273 are slightly affected by pile up, by considering the single and double event ratios, 
though the data curve almost exactly follows the model curve. For these observations removing the central region of the source 
did not improve the ratios. Hence, we ignore pile up for those observations.

\subsection{Excess Variance}

The excess variance, $\sigma_{XS}$ and fractional rms variability amplitude, F$_{var}$, (e.g. Edelson et al. 2002) are commonly 
used parameters to quantify variability. Excess variance gives the intrinsic variance of the 
source by removing the variance due to measurement errors in each individual flux measurement. The fractional rms variability 
gives the average variability amplitude with respect to the mean flux of a source. The uncertainty on F$_{var}$ has been 
calculated using Vaughan et al. (2003; see Eq. (B2) there). We have given the detailed description of the method in Kalita et 
al. (2015).

The excess variance of  $n$ number of flux measurements $x_{i}$,  at times $t_{i}$, with  measurement errors $\sigma_{err,i}$, 
is calculated as
\begin{equation}
\sigma_{XS}^{2} = S^{2} - \overline{\sigma_{err}^2},
\end{equation}
where $\overline{\sigma_{err}^{2}}$ is the mean square error and $S^2$ is sample variance

then fractional rms variability can be calculated as follows, 

\begin{equation}
F_{var} = \sqrt{ \frac{ S^{2} -
\overline{\sigma_{err}^{2}}}{\bar{x}^{2}}}
\end{equation}

Here, $\bar{x}$, represents mean count of the source.

The uncertainty on $F_{var}$ has been calculated here, using the equation (B2) in Vaughan et al. (2003) for $n$ number of events, 
which is  given by

\begin{equation}
(F_{var})_{err}=\sqrt{ \left\{ \sqrt{\frac{1}{2n}} \frac{ \overline{\sigma_{err}^{2}}
}{\bar{x}^{2}F_{var} } \right\}^{2}+\left\{ \sqrt{\frac{\overline{\sigma_{err}^{2}}}{n}}\frac{1}{\bar{x}}\right\}^{2}}
\end{equation}

\section{Results}

In Table 2, we report the detailed analysis result of all the 50 LCs of 10 LSPs and ISPs. 
 The sample of LSPs studied here is the collection of archival observations of all LSPs with XMM-Newton since its launch. 
To avoid any type of observational bias, we have taken all the observations of LSPs inspite of  
their flux states which can be low-state, pre-outburst, post-outburst or outburst.  The quality of detection of each observation 
by XMM-Newton is represented by its signal-to-noise ratio (SNR), listed in the last column of the Table 2. Different values of SNR 
represents different detection limit. If for particular pointing, SNR $\geq$ 5 is called strong detection, 3 $<$ SNR $<$ 5, is 
called border line 
detection, and $\textless$ 3 is considered as not significant detection. Below we report the results of IDV detection in
individual LSPs and ISPs.

\subsection{TXS 0106+612}

In SIMBAD this source is tagged as ``possible blazar". But MOJAVE collaboration classified it as LSP. This source 
is also known as Fermi J0109+6133 (J0109+6134), or GT 0106+613. The source was first discovered by Gregory $\&$ Taylor (1981) 
in a radio survey of the Galactic plane.
During the XMM--Newton's pointing in 2010 February, the source was in flaring state detected by Fermi Large Area Telescope (LAT) 
(Vandenbroucke et al. 2010). 
AGILE and Swift satellites also partly covered this 
event in gamma-ray and X-rays bands, respectively. No significant variation was detected with Swift satellite in X-ray band 
(Vandenbroucke et al. 2010). 

The single observation studied here show $F_{var}$ $\sim$ 11 $\%$ with SNR $\sim$ 1.19 which is below detection limit. Hence, 
we can not claim that the source has shown genuine IDV. 

\subsection{PKS 0235+164}

PKS 0235+164 is a  BL Lac object and well studied in the entire EM spectrum. In optical band, the source was found to 
be very active and has shown IDV 
(Romero et al. 2000, Gupta et al. 2008). 
Using 25 years (1975--2000) of radio and optical data, Raiteri et al. (2001) predicted that the blazar should show a 
possible correlated radio and optical outburst in 2004. But the source did not show radio/optical outburst in 2004 as 
predicted, and significant 
variation in X-ray band was also not found by the help of 
contemporary observations carried out by VLBI, ground based optical telescopes and XMM--Newton (Raiteri et al. 2005). 
In the observing period of 2000--2005 (with Chandra and XMM--Newton), the source was found in X-ray bright state in 2002. 
Except in 2002, the source was in its faint state, with hard spectrum (Raiteri et al. 2006). Flaring in the source in 
different EM bands from radio to 
gamma-ray including X-ray is reported by Agudo et al. (2011), and have shown correlation in X-ray, optical, millimeter 
and centimeter bands. 

We analyzed four pointings of XMM-Newton of PKS 0235$+$164 and found significant IDV in one in 2002 when
the source was in outburst state. Other observations were done in year 2004, 2005 when the source was in faint 
state with SNR less than 5 i.e. below the genuine detection limit. Hence, no significant IDV can be claimed.

\subsection{PKS 0426-380}

The blazar shows high variation in GeV band on long time scales (Neronov et al. 2015) and the source flux enhanced by a factor 
of 2--3 during its flaring activity. This is the most distant HE gamma-ray emitting source so far which is reported in Tanaka 
et al. (2013) detected by Fermi LAT instrument in January 2013. During this period the source was in flaring state in gamma-ray.

This source has single observation with XMM--Newton taken in 2010. The observation has SNR $<$ 2 i.e. the source was below the 
detection limit and no genuine IDV can be reported.

\subsection{PKS 0528+134}

PKS 0528+134 is one of the most distant (z = 2.07) and bright gamma-ray blazar. 
The 2009 XMM-Newton observation of the source was part of a multi-wavelength campaign during its quiescent state (Palma 
et al. 2011). They reported absence of flux and spectral variability in the gamma-ray, X-ray and radio bands on IDV 
timescales, but found significant flux variation on IDV timescales in optical bands. They have also reported STV of 
moderate strength in the X-ray and radio regime on 1-2 week timescales. The simultaneous multi-wavelength SED shows 
that the bolometric luminosity is dominated by gamma-ray emission. As usual the low-energy SED peak lies within infrared/optical 
bands, and the high-energy SED hump peaks at MeV -- GeV energies.

We analyzed single XMM--Newton observation of the blazar taken in 2009 when it was in low state and we did not find any
significant detection of IDV in this particular pointing.

\subsection{PKS 0537-286}

PKS 0537-286, the high redshift (z = 3.104) FSRQ was studied by Reeves et al. (2001) with XMM-Newton. They reported 
that the source is extremely luminous in X-ray band. In their spectral study, they found that in contrast with typical 
AGNs, the  radio to X-ray SED of the source shows dominance of X-ray power over other energy bands, which is explained 
by inverse Compton (IC) emission. Optical 
variability with different instruments and marginal 
X-ray variability on IDV timescale with Swift/XRT was reported by Bottacini et al. (2010). 
Sambruna et al. (2007) reported that there were no soft and hard X-ray flux variability in the source with different XRT 
and BAT pointings within the period 2005-2007.

We analyzed two observations for the blazar taken with XMM-Newton in 2000 and 2005. We did not find any significant 
IDV on both the occasions. No earlier information about the source flux based on other EM bands at the time of XMM-Newton 
observations is available.

\subsection{S5 0716+714}

The BL Lac S5 0716+714 is classified as ISP blazar (e.g. Giommi et al.1999; Giommi et al. 2008; and references therein).
It has been prime interest 
of IDV study in optical bands and variability on other time scales has been reported frequently for the blazar (e.g. Heidt 
$\&$ Wagner 1996; Gupta et al. 2008; and references therein).
During the XMM-Newton observation, the blazar S5 0716+714 was in high state with multiple flares (Zhang 2010). 
The temporal and spectral variability study of the source was done in detail by Zhang (2010), where they reported that 
within the 74 ks exposure, 
the soft X-ray (0.5--0.75 keV) flux varies by a factor of $\sim$ 4, which is much stronger as compared to the hard X-ray 
(3--10 keV) variations with $F_{var}$ values 32.03$\pm$2.12 $\%$ and  7.62$\pm$1.5 $\%$, respectively. 
XMM-Newton observations of S5 0716+714 have shown variation by more than a factor of 3 on timescale of hours (e.g.
Ferrero et al. 2006). 
%X-ray flux variation of S5 0716+714 with XMM-Newton observations by more than a factor of 3 on timescale of hours (e.g.  
%Ferrero et al. 2006). 
They reported that the variability amplitude in soft (0.5--0.75 keV) and hard bands (3--10 keV)
are 0.40$\pm$0.03 and 0.27$\pm$0.01, respectively and synchrotron emission was dominated during the flaring activity.  This
is an IBL in which first hump peaks at NIR/optical region, and X-ray will fall below the peak in which hard X-ray will have 
comparatively retarded effect than soft X-ray. It will cause here high amplitude variation in soft X-ray compare to hard X-ray.   

We analyzed only one observation of the source with XMM-Newton which is taken during its high state and found strong IDV in it. 

\subsection{4C 71.07}

S5 0836+710 (4C 71.07) is a high redshift FSRQ. 
 With Swift observations, X-ray variability was found on a time scale 
of one month in the energy range 15-150 keV (Sambruna et al. 2007). The source has shown the most luminous gamma-ray 
flare from any blazar till date (during Oct-Dec in 2011) in 0.1--300 GeV during its outburst state (Paliya 2015), 
where IDV is also reported. Multiple X-ray and gamma-ray flares are reported by the author and it was noticed that the 
optical-UV flux do not show any significant variation since these bands are dominated by accretion disk emission. Different 
satellite's (including NuSTAR, SWIFT) observations of the source reported that IDV in X-ray band is absent but significant 
STV in different X-ray energy ranges (Ghisellini et al. 2010; Akyuz et al. 2013; Tagliaferri et al. 2015). 
Synchrotron hump of SED lies in sub-millimeter range (Paliya 2015). 
The XMM-Newton observation of the source was reported by Foschini et al. (2006) for the first time to study spectral 
behaviour of gamma-ray loud AGN sample. The spectra of the source gives best fit with simple power law with galactic 
absorption.

We analyzed one observation of this source taken by XMM-Newton in 2001 and did not find any significant detection 
of IDV during this pointing. 

\subsection{OJ 287} 
OJ 287 is one of the most studied BL Lac object in the complete EM spectrum. It is claimed that there is binary super 
massive black hole (SMBH) system (Sillanap$\ddot{a}\ddot{a}$ et al. 1988). 
 XMM-Newton pointings of the source in 2005 are organized in correspondence with detected optical 
flare and expected optical outburst (Ciprini 2005; Ciprini et al. 2007). The source was studied in X-ray bands 
during its optical 
bright states by Massaro et al. (2003), where they have found low X-ray flux and the SED synchrotron peak falls in IR band. 
In 2005 November the source was in outburst state in optical band reported by Valtonen et al. (2006).
Again an optical outburst occurred in September 2007 (Valtonen et al. 2009). The source was reported as in pre/post outburst
 state in optical band during October 2006 to January 2007 by (Gupta et al. 2008). 
Another optical outburst in the object occurred during October 2007 to January 2008 (Dai et al. 2011) and During XMM-Newton's 
pointing in 2008 the object was in post-outburst state (Villforth et al. 2010). 
 A major optical outburst in the blazar is reported in 2012--2013 (Carnerero et al. 2015).

Here, we have analyzed 5 XMM-Newton pointings of the source and none of them show genuine IDV.

\subsection{3C 273}

3C 273 is the first quasar discovered (Schmidt 1963). 
The source has been studied extensively in the entire EM bands. 
In X-ray bands the source show soft excess below $\sim$ 1-2 keV (Turner et al. 1985; Page et al. 2004; Chernyakova 
et al. 2007), which has been interpreted as result of combination of Seyfert and blazar like emission in this band 
(e.g., Soldi et al. 2008; Pietrini \& Torricelli-Ciamponi 2008). 
3C 273 is very bright and nearby FSRQ, but IDV study for the source is rare in X-rays. 
We have searched for IDV with XMM-Newton pointings from 2000--2012, but no significant IDV has been found (Kalita et al. 
2015). 
The source was in low state during the June 2003 XMM-Newton observation, and it reached a historically softest state in X-ray 
band (Chernyakova et al. 2007). 
RXTE-PCA observed the source from March 2005 to December 2011, during which several consecutive X-ray flares were observed 
corresponding high and low state (Esposito et al. 2015). 

Since the launch of XMM-Newton in 2000, most extensively this blazar is pointed by XMM-Newton. In total we have 25 pointing
of the blazar for IDV study and genuine IDV is not detected in any pointing (Kalita et al. 2015).   

\subsection{3C 279}

3C 279 is one of the most studied FSRQ and classified as a LSP (Ackermann et al. 2011). 
A gamma-ray flare was reported by Hayashida et al. (2012), which has a optical counterpart with 10 days delay, 
and they reported two X-ray flares with a 90 days separation, which was not related with the gamma-ray/optical flares. 
The LSP shows weak variability in X-ray flux on IDV time scales with NuSTAR observations and shows significant 
spectral variation in the Swift satellite data reported in Hayashida et al. (2015). Chatterjee et al. (2008) studied 
the source in optical, X-ray, radio band and found that the variability amplitude increases with timescales and the variations 
in X-ray and optical band are well correlated. 
The source was in optical outburst state 
in 2007 in optical (Gupta et al. 2008), with absent of IDV but STV was reported. The source has been studied less frequently 
in X-ray bands. The source have shown rapid flare in X-ray bands with a duration of 20 days in April 2009 and a second X-ray flare was observed
after 3 months of the first flare (Abdo et al. 2010b). 

We analyzed two pointings of this source taken with XMM-Newton in 2009 and 2011 but did not detect any significant IDV in these 
observations.

\subsection{BL Lac}

The prototype of its class, BL Lac is one of the most extensively studied blazar in the entire EM spectrum. Flux variability 
on diverse time scales in the optical bands have been reported (Massaro et al. 1998; Clements $\&$ Carini 2001; Villata et al. 
2002; Zhai \& Wei 2012; Gaur et al. 2015; and references therein). Several multi-wavelength campaigns were triggered in the 
past to study in detail the properties and behaviour of the object (e.g. Villata et al. 2002; Papadakis et al. 2007; Raiteri 
et al. 2009; Jorstad et al. 2010; Marscher et al. 2010; Raiteri et al. 2013; and references therein). Using BeppoSAX observations
genuine X-ray IDV detection in the blazar is reported (Ravasio et al. 2002, 2003; B{\"o}ttcher et al. 2003). Raiteri et al. (2009) 
analyzed three XMM-Newton observations during 2007--2008 in comparatively low state and found that the SED synchrotron peak 
falls in the near IR region. In the low state of the blazar, the multi-frequency campaign during second half of 2008, was 
reported by Abdo et al. (2011). Moderate IDV flux variation ($\sim$ 4--7$\%$) was reported on hour time scales in X-ray bands 
with XMM-Newton observations during 2008 (Raiteri et al. 2010) but fast STV was detected in Swift X-ray data by Raiteri et al. (2010).

We analyzed three observations of this source in 2007--2008 and in any of these observations, we did not find significant 
IDV in any of the observation. 

\subsection{3C 454.3}

3C 454.3 is a well studied FSRQ in the entire EM spectrum. The source has been flaring and showing outburst from radio 
to gamma-ray at different epochs (e.g., Vercellone et al. 2008; Ghisellini et al. 2007; Giommi et al. 2006; Villata et al. 
2006; Pian et al. 2006; Fuhrmann et al. 2006, Foschini et al. 2010; Bachev et al. 2011; Jorstad et al. 2013; Sasada et al. 
2014; Pacciani et al. 2014; Kohler $\&$ Nalewajko 2015). 
In a multi-wavelength monitoring campaign of 3C 454.3 in 2008, Bonning et al. (2009) found that the source was highly variable 
in NIR, optical, UV and gamma-rays, but in X-ray band it was not variable. Multi-peak outburst in optical band was observed in 
July - August 2007 and November 2007 - February 2008, where IDV was detected in several episodes, with correlated X-ray and optical 
flux (Raiteri et al. 2008).
Even in outburst states the SED synchrotron peak lies within the IR region (Giommi et al. 2006; Fuhrmann et al. 2006)
The source was observed by XMM-Newton during 2006 in the post outburst phase (Raiteri et al. 2007).

We analyzed four observations of this source taken in 2006 and 2007 in the post outburst state. We did not detect any significant 
IDV during these observations. 

\begin{table*}
{\bf Table 2.} X$-$ray variability parameters of Low Energy Peaked Blazars
\small
\begin{tabular}{lccccccc} \hline 

Blazar Name   & Obs. Date   &Obs.ID         & Variance    & $\sigma_{XS}^2$ & F$_{var}$ (percent) & SNR$^{a}$ &$F^{b}$ \\
              &             &               &             &                 &                        &      &ergs $cm^{-2}s^{-1}$ \\\hline

TXS 0106$+$612   & 2010.02.09  &0652410201 	  & 0.002       & 0.001   & 11.394$\pm$6.539    &1.194 $^{\star}$&  -12.097   \\
PKS 0235$+$164   & 2002.02.10  &0110990101 	  & 0.214       & 0.169   & 11.600$\pm$0.495    &24.322          & -10.842\\
                 & 2004.01.18  &0206740101 	  & 0.012       & 0.001   & ~3.173$\pm$5.321    &2.752$^{\star}$ &  -11.549\\
                 & 2004.08.02  &0206740501 	  & 0.014	& 0.001	  & ~5.187$\pm$2.204    &4.254$^{\star\star}$ & -11.534\\
                 & 2005.01.28  &0206740701 	  & 0.005	& 0.001	  & ~9.025$\pm$3.836    &1.924$^{\star}$ & -11.880\\
PKS 0426$-$380   & 2012.02.11  &0674330201 	  & 0.004	& 0.001	  & ~2.606$\pm$11.145   &1.927$^{\star}$ &  -12.131      \\
PKS 0528$+$134   & 2009.09.11  &0600121601 	  & 0.004	& 0.001	  & ~2.952$\pm$8.832    &1.973$^{\star}$ & -11.838\\
PKS 0537$-$286   & 2000.03.19  &0114090101 	  & 0.014       & 0.002	  & ~4.723$\pm$1.994    &6.402           & -11.622\\
                 & 2005.03.20  &0206350101 	  & 0.036       & 0.005	  & ~4.771$\pm$2.038    &9.603           & -11.413\\
S5 0716$+$714    & 2007.09.24  &0502271401 	  & 1.423	& 1.338	  & 23.007$\pm$0.236    &12.414          & -10.856\\
4C 71.07         & 2001.04.12  &0112620101 	  & 0.130	& 0.026	  & ~1.345$\pm$0.602    &60.105          & -10.313\\
OJ 287           & 2005.04.12  &0300480201 	  & 0.018     	& 0.002   & ~4.036$\pm$2.861    &8.941           & -11.466\\
                 & 2005.11.03  &0300480301 	  & 0.016	& 0.001	  & ~2.097$\pm$2.300    &7.825           & -11.477\\
                 & 2006.11.17  &0401060201 	  & 0.012	& 0.001	  & ~1.889$\pm$2.289    &6.940           & -11.563\\
                 & 2008.04.22  &0502630201 	  & 0.013       & 0.001	  & ~2.327$\pm$1.990    &5.346           & -11.566\\
                 & 2011.10.15  &0679380701 	  & 0.040	& 0.001	  & ~0.949$\pm$1.908    &23.025          & -11.035\\
3C 273           &2000.06.13   &0126700301  	& 0.764	& 0.105	          & 0.689$\pm$0.148     &221.728         &   -9.899 \\
                 &2000.06.15   &0126700601  	& 0.633	& 0.005	          & 0.162$\pm$0.847     &213.119         &   -9.914\\
                 &2000.06.15   &0126700701 	& 0.545  & 0.087	  & 0.666$\pm$0.238     &183.692         &   -9.926        \\
                 &2000.06.17   &0126700801  	& 0.774  & 0.152	  & 0.881$\pm$0.140     &162.498         &   -9.925  \\
                 &2001.06.13   &0136550101  	& 0.303	& 0.029	          & 0.795$\pm$0.206     &77.883          &  -10.237        \\
                 &2001.12.16   &0112770101  	& 0.816	& 0.193	          & 0.610$\pm$0.400     &199.081         &   -9.718 \\
                 &2001.12.22   &0112770201  	& 1.041	& 0.057           & 0.345$\pm$0.665     &210.513         &   -9.733\\
                 &2002.07.07   &0112770601  	& 0.563	& 0.202	          & 0.833$\pm$0.421     &106.273         &   -9.843\\
                 &2002.12.17   &0112770801 	& 0.952	& 0.132	          & 0.467$\pm$0.457     &169.196         &   -9.686\\
                 &2003.01.05   &0112770701  	& 0.787	& 0.126	          & 0.545$\pm$0.474     &223.767         &  -9.760\\
                 &2003.01.05   &0136550501 	& 0.895	& 0.006	          & 0.124$\pm$1.475     &225.361         &   -9.770 \\
                 &2003.06.18   &0112771001  	& 1.156	& 0.023	          & 0.187$\pm$1.007     &285.664         &   -9.671 \\                 
                 &2003.07.07   &0159960101 	& 0.786	& 0.456	          & 2.593$\pm$0.114     &109.873         &  -10.159        \\
                 &2003.07.08   &0112770501 	& 0.711	& 0.373	          & 0.865$\pm$0.272     &219.411         &   -9.726\\
                 &2003.12.14   &0112771101  	& 0.689	& 0.076	          & 0.518$\pm$0.462     &193.369         &   -9.848  \\
                 &2004.06.30   &0136550801  	& 0.848	& 0.201	          & 0.992$\pm$0.221     &119.784         &   -9.915\\
                 &2005.07.10   &0136551001 	& 0.601	& 0.103           & 0.651$\pm$0.223     &242.392         &   -9.875 \\
                 &2007.01.12   &0414190101 	& 0.332	& 0.061	          & 1.151$\pm$0.172     & 90.373         &  -10.240\\
                 &2007.06.25   &0414190301  	& 0.624	& 0.031	          & 0.361$\pm$0.372     &201.652         &   -9.909 \\
                 &2007.12.08   &0414190401 	& 1.444	& 0.963	          & 2.571$\pm$0.121     &108.421         &   -9.989 \\
                 &2008.12.09   &0414190501  	& 0.562	& 0.281	          & 2.467$\pm$0.162     & 62.124         &  -10.241        \\
                 &2009.12.20   &0414190601 	& 0.476	& 0.154	          & 1.612$\pm$0.202     & 80.801         &  -10.189 \\
                 &2010.12.10   &0414190701 	& 0.304	& 0.073	          & 1.481$\pm$0.237     & 65.194         &  -10.313\\
                 &2011.12.12   &0414190801 	& 1.582	& 0.911	          & 2.009$\pm$0.101     &194.326         &   -9.899         \\
                 &2012.07.16   &0414191001 	& 0.693	& 0.178	          & 1.023$\pm$0.280     &207.683         &   -9.960 \\
3C 279           & 2009.01.21  &0556030101 	  & 0.056	& 0.007	  & ~1.740$\pm$0.943  &29.397          &  -10.888\\
                  & 2011.01.18  &0651610101 	  & 0.070	& 0.001	  & ~0.640$\pm$1.011  &16.678          &  -10.876\\
BL Lac            & 2007.07.10  &0501660201 	  & 0.069	& 0.008	  & ~1.931$\pm$1.055  &16.703          & -10.699\\
                  & 2007.12.05  &0501660301 	  & 0.070	& 0.010	  & ~2.750$\pm$0.948  &21.135          &  -10.803\\
                  & 2008.01.08  &0501660401 	  & 0.057       & 0.005	  & ~1.939$\pm$1.156  &21.436          &  -10.792\\
3C 454.3          & 2006.07.02  &0401700201 	  & 0.054       & 0.007	  & ~3.001$\pm$1.226  &19.269          &  -11.069\\
                  & 2007.05.23  &0401700401 	  & 0.240	& 0.055	  & ~2.090$\pm$1.138  &27.748          & -10.472\\
                  & 2006.12.18  &0401700501 	  & 0.062	& 0.003	  & ~1.437$\pm$2.003  &17.443          &  -10.939\\
                  & 2007.05.31  &0401700601 	  & 0.158       & 0.022	  & ~1.370$\pm$0.513  &34.568          & -10.486\\\hline
\end{tabular}     \\
%\textbf
$^{a}$ Signal to noise ratio of the data, $^{\star}$ not significant detection, $^{\star \star}$ border line detection,\\
$^{b}$ Log of flux in the energy range 0.3--10 keV, calculated using the tool ``WebPIMMS" provided by NASA's HEASARC considering a power law spectra with galactic absorption.
\end{table*}

\section{Discussion and Conclusion}

We searched for X--ray IDV in the LCs of almost all the LSPs observed by XMM--Newton since its launch. The details of the observation 
log is given in Table 1 which gives the data of 9 LSPs and 2 ISPs with total 25 LCs. All the LCs are plotted in Fig. 1. There were 25
LCs of LSP 3C 273 which are already presented in Table 1 and Fig. 1 of Kalita et al. (2015). In Table 2, we report the results of 
variability parameters of all 50 LCs which are calculated using excess variance. It is clear from Table 2 that in only two pointings, \
one of the ISP S5 0716+714 and another one of LSP PKS 0235+164, we found significant IDV with variability amplitude of 23 and 11.6\%, 
respectively. Since, S5 0716+714 is observed in outburst state and is an ISP, it could be expected that the synchrotron peak of this 
blazar reaches up to soft X-ray regime. Similarly, PKS 0235+164 is also found to shift its peak up to soft X-ray region in its outburst 
state (Madejsi et al. 1996). Hence, in these two occasions, we expect to get the variability in X-ray band. In other pointings, we did 
not find any significant variability. 
Most of the sources listed in Table 1. are well known LSPs and are extensively monitored in optical bands and they have shown high 
duty cycles in optical bands up to $\sim$ 70--80\% (Gupta et al. 2008, 2012; Goyal et al. 2012; Agarwal et al. 2015; Gaur et al. 2015 
and references therein). 

Similarly, IDV of HSPs in X-ray band is well studied and are highly variable in these bands (Lachowicz et al. 2009; Gaur et al. 2010; 
Kalita et al. 2015; and references therein) but not very extensively studied in optical bands. Till now, we put an extensive effort to 
observe the optical IDV of HSPs and found these sources to be less variable in these bands. 

The difference in the multi-frequency spectral properties of HSPs and LSPs requires a systematic change of intrinsic
physical parameters such as magnetic field, jet size, maximum electron energy and it is investigated by Sambruna et al. (1996)
that the change is in the sense that HSPs have higher magnetic fields/electron energies and smaller sizes as compared to LSPs.
All the above factors lead to the difference in the cut off energies of LSPs and HSPs and hence have a more retarded 
effect on the X-ray variability of LSPs. ISPs lies in between these two classes and is difficult to mark the exact boundaries 
as it depends on the state of the source i.e whether it is in quiescent/outburst state. 
The variations in the acceleration efficiency of the relativistic electrons near the synchrotron hump could arise from the changes 
in the local number density of the most energetic electrons or the strength of the
localized magnetic fields. Near the peak of the SED, acceleration processes dominates and produces the higher energy electrons
while the lower energy electrons are available for the emission below the SED peaks. Below the SED peak, probably cooling processes
dominate which involve mainly inverse Compton for LSPs in general 
%FSRQs as they are dominated by the external seed photons 
(Joshi et al. 2014).  
Since, the most excepted model for the intra-day variability involves magneto-hydrodynamic instabilities; presence of turbulence behind or in the vicinity 
of the shock (Marscher, Gear \& Travis 1992); hence one can expect that the X-ray 
variability would be more pronounced for HSPs as compared to LSPs which is confirmed from our observations.  
These differences can lead to the apparent dichotomy between these two classes of blazars however we need statistically more 
observations to firmly conclude our findings.

\section*{ACKNOWLEDGMENTS}

We thank the referee for important comments and suggestions which helped us to improve the manuscript. This research is based 
on observations taken with XMM--Newton, an ESA science mission with instruments and contributions directly funded by 
ESA Member Sates and NASA. ACG is partially supported by the Chinese Academy of Sciences (CAS) Visiting Fellowship for Researchers 
from Developing Countries (grant no. 2014FFJA0004) and CAS President's International Fellowship Initiative (PIFI) (grant no. 
2016VMB073. NK acknowledge the Department of Science \& Technology (DST), India, for 
supporting this work with grants under the Women Scientist scheme--A (WOS--A). 
HG is sponsored by the Chinese Academy of Sciences (CAS) Visiting Fellowship for Researchers from Developing
Countries, CAS Presidents International Fellowship Initiative (grant no. 2014FFJB0005), supported by the NSFC Research Fund for
International Young Scientists (grant No. 11450110398) and supported by a Special Financial Grant from the China Postdoctoral
Science Foundation (grant No. 2016T90393).

\clearpage

\end{document}